\pdfoutput=1

\RequirePackage{etoolbox}
\csdef{input@path}{
 {img/}
}

\documentclass{article}

\usepackage{arxiv}

\usepackage[utf8]{inputenc} 
\usepackage[T1]{fontenc}    
\usepackage{hyperref}       
\usepackage{url}            
\usepackage{booktabs}       
\usepackage{amsfonts}       
\usepackage{nicefrac}       
\usepackage{microtype}      
\usepackage{lipsum}
\usepackage{amsthm}
\usepackage{amsmath}
\usepackage{amssymb}
\usepackage{titlesec}
\usepackage{gensymb}
\usepackage[linesnumbered,ruled,vlined]{algorithm2e}
\usepackage[nomarkers,figuresonly]{endfloat}
\usepackage{graphicx}
\usepackage{mathtools}

\title{Heteroscedastic Gaussian Process Regression on the Alkenone over Sea Surface Temperatures}

\author{
  Taehee Lee \\
  Division of Applied Mathematics\\
  Brown University\\
  Rhode Island, USA \\
  \texttt{taehee\_lee@brown.edu} \\
  \And
  Charles E. Lawrence \\
  Division of Applied Mathematics\\
  Brown University\\
  Rhode Island, USA \\
  \texttt{charles\_lawrence@brown.edu} \\
}

\begin{document}
\maketitle

\begin{abstract}
To restore the historical sea surface temperatures (SSTs) better, it is important to construct a good calibration model for the associated proxies. In this paper, we introduce a new model for alkenone (${\rm{U}}_{37}^{\rm{K}'}$) based on the heteroscedastic Gaussian process (GP) regression method. Our nonparametric approach not only deals with the variable pattern of noises over SSTs but also contains a Bayesian method of classifying potential outliers.
\end{abstract}

\section{Introduction}\label{sec1}

The alkenone is a widely used proxy for inferring sea surface temperatures (SSTs) in paleoceanography. Haplophytes make long-chain ketone lipids (alkenones) with 37 carbons changing in response to water temperature. Let ${\rm{U}}_{37}^{\rm{K}'}$ be the relative unsaturation index of these compounds as \cite{TIERNEY2018}:
\begin{equation}
{\rm{U}}_{37}^{\rm{K}'} = \frac{{\rm{C}}_{37:2}}{{\rm{C}}_{37:2}+{\rm{C}}_{37:3}} \label{equ0}
\end{equation}

Because SSTs have other sources of inferences, what we are interested in is the distribution of ${\rm{U}}_{37}^{\rm{K}'}$ given SST, not the reverse: once the prior information (based on the latitude, for example) of SST is organized as a prior distribution and the distribution as a likelihood, it is possible to integrate those information by computing the posterior distribution of SST given the observed ${\rm{U}}_{37}^{\rm{K}'}$ by the Bayes' rule. Also, if SSTs are somehow correlated (with respect to their spatial pattern, for example) one another, the given proxies can be exploited better than a set of individual posteriors with an associated graphical model.

\cite{TIERNEY2018} has lead the way in the application of Bayesian statistics in paleoclimatology. The paper approaches to the problem with a Bayeian B-spline regression model (BAYSPLINE) on ${\rm{U}}_{37}^{\rm{K}'}$ data as well as considering their seasonality, and it shows improved performance to extant methods. However, the model lacks of a concrete rule of deciding the number of basis splines or their orders, and does not deal with heteroscedastic noises over SSTs neatly.

For the last decade, models based on the neural networks (NN) have been rapidly emerging and soon overwhelming all research fields requiring machine learning, including climatology, for the outstanding performance. Though some theoretical works such as \cite{CYBENKO1989,HORNIK1991,LU2017} guarantee the effectiveness of NNs in some senses, it remains as a potential problem in practice that parameters to be learned are often too many compared to the amount of available data. Bayesian statistics can give a possible solution of this problem: its philosophy that parameters are random variables allows to marginalize all such unknown values out to get the posterior predictive distributions.

In this point of view, among various nonparametric regression models, Gaussian processes (GP) \cite{Rasmussen2005} have some advantages which make them distinctive from all the others. Besides their explicit predictive posterior distributions, with some specific kernels corresponding to the choice of activation functions, it becomes a marginalized version of associated NN models: details can be found in \cite{WILLIAMS1998,CHO2010}. The point is that only a very few hyperparameters (for example, only three hyperparameters are needed in the squared-exponential kernel) are now controlling the GP regression models, and let data explain themselves to make them free from the structure.

Though tuning hyperparameters can be done efficiently in the homoscedastic (i.e., noises are assumed to have a constant variance.) GP models \cite{Rasmussen2005}, heteroscedastic GP models are not yet clear to learn. While \cite{Kersting2007} tries to model logarithms of variances from empirical variances based on another GP regression, \cite{Lazaro-Gredilla2011} adapts a variational method which cannot avoid breaking up the completeness of the models. In this paper we suggest a heteroscedastic GP model which is not only intuitive but also easy to learn and apply it to constructing a new calibration for ${\rm{U}}_{37}^{\rm{K}'}$.

There is one golden rule: every Gaussian model is sensitive to outliers in learning. Also, we should be aware of the model misspecification for the robustness, unless the model is guaranteed to follow a certain distribution based on the underlying theoretical arguments. GP models cannot be free from these limitations. Student's t-processes \cite{Shah2014} can be an alternative, but they do not have explicit posterior predictive distributions if noises are also considered. Instead, classifying and excluding outliers in learning seems to be a better idea. We also introduce a Bayesian method of classifying and treating outliers automatically in the learning procedure.

\section{Method}\label{sec2}

Let $\mathcal{X}={\left\{x_n\right\}}_{n=1}^{\rm{N}}$ and $\mathcal{Y}={\left\{y_n\right\}}_{n=1}^{\rm{N}}$ be the SSTs and ${\rm{U}}_{37}^{\rm{K}'}$ observations, respectively. What to construct is the regression model which returns the distribution $p \left( y \middle| x,\mathcal{X},\mathcal{Y} \right)$ of ${\rm{U}}_{37}^{\rm{K}'}$, $y$, at an arbitrary query SST, $x$. Let $\beta$ be the regression function of $\mathcal{Y}$ at $\mathcal{X}$. Then, what we assume are the following:
\begin{equation}
p \left( \mathcal{Y} \middle| \beta , \mathcal{X} \right) = \mathcal{N} \left( \mathcal{Y} \middle| \beta , \Lambda ( \mathcal{X} ) \right) \label{equ1}
\end{equation}
\begin{equation}
p \left( \beta \middle| \mathcal{X} \right) = \mathcal{GP} \left( \beta \middle| \vec{0} , \mathbb{K} \left( \mathcal{F} (\mathcal{X}) , \mathcal{F} (\mathcal{X}) \right) + {\xi}^{2} \mathbb{I} \right) \label{equ2}
\end{equation}
, where $\Lambda$ is a function returning variances of ${\rm{U}}_{37}^{\rm{K}'}$ observations in the form of a diagonal matrix at the query SSTs given their regression vector, $\mathcal{F}$ is a feature function, and $\mathbb{K}$ is the GP kernel function defined as follows: for a kernel function $k$,
\begin{equation}
\mathbb{K} ({\mathcal{X}}^{(1)},{\mathcal{X}}^{(2)}) \triangleq { \left[ k \left( \mathcal{F} ( {\mathcal{X}}_{m}^{(1)} ) , \mathcal{F} ( {\mathcal{X}}_{n}^{(2)} ) \right)  \right] }_{m,n} \label{equ3-2}
\end{equation}
, where $\rm{A} \triangleq \rm{B}$ means that A is defined by B.

Because feature functions may not be injective, the term ${\xi}^{2} \mathbb{I}$ is inevitable to guarantee (\ref{equ2}) to be well-defined by making it nondegenerate.

Then, for a scalar regression $b$ of $y$ at $x$, the regression distribution $p \left( y \middle| x , \mathcal{X} , \mathcal{Y} \right)$ is derived as follows:
\begin{equation}
p \left( y \middle| x , \mathcal{X} , \mathcal{Y} \right) = \int{ p \left( y \middle| b , x \right) \int{ p \left( b \middle| \beta , x , \mathcal{X} \right) p \left( \beta \middle| \mathcal{X} , \mathcal{Y} \right) } d\beta db  }
\label{equ4}
\end{equation}

As consistent with (\ref{equ1}), we have the likelihood of $y$ given $b$ and $x$ as a Gaussian $p \left( y \middle| b , x \right) = \mathcal{N} \left( y \middle| b , \Lambda (x) \right)$.

Because the density of regression vector given SSTs is assumed to be a GP, we have the following extension:
\begin{equation}
p \left( \beta , b \middle| \mathcal{X}, x \right) = \mathcal{N} \left( \begin{pmatrix} \beta \\ b \end{pmatrix} \middle| 0 , \begin{pmatrix} \mathbb{K}_{11} + {\xi}^{2} \mathbb{I} & \mathbb{K}_{12} \\ \mathbb{K}_{21} & \mathbb{K}_{22} + {\xi}^{2} \end{pmatrix} \right) 
\label{equ6}
\end{equation}
, $\mathbb{K}_{11} \triangleq \mathbb{K}\left( \mathcal{F} \left( \mathcal{X} \right) , \mathcal{F} \left( \mathcal{X} \right) \right)$, $\mathbb{K}_{22} \triangleq \mathbb{K}\left(\mathcal{F}\left(x\right),\mathcal{F}\left(x\right)\right)$, $\mathbb{K}_{12} \triangleq \mathbb{K} \left( \mathcal{F} \left( \mathcal{X} \right) , \mathcal{F} \left( x \right) \right)$ and $\mathbb{K}_{21} \triangleq \mathbb{K}\left(\mathcal{F}\left(x\right),\mathcal{F}\left(\mathcal{X}\right)\right)$ are the abbreviations. Thus, the conditional distribution of $b$ given $\beta$, $\mathcal{X}$ and $x$ can be computed explicitly as a Gaussian.

Finally, the posterior distribution of $\beta$ given $\mathcal{X}$ and $\mathcal{Y}$ is derived from the likelihood (\ref{equ1}) and prior (\ref{equ2}), which is also a Gaussian.

I.e., three terms in (\ref{equ4}) are all Gaussian, so $p\left(y\middle| x,\mathcal{X},\mathcal{Y}\right)$ can be computed analytically as follows: note that $\left( \mathbb{K}_{11} + \xi^2 \mathbb{I} + \Lambda \left( \mathcal{X} \right) \right)^{-1}$ is not a function of $x$ or $y$, so only one matrix inversion is required to compute the following $\mu$ and $\nu$.
\begin{equation}
p \left( y \middle| x , \mathcal{X} , \mathcal{Y} \right) =  \mathcal{N} \left( y \middle| {\mu} \left( x , \mathcal{X} , \mathcal{Y} \right) , \Lambda \left( x \right) + {\nu} \left( x , \mathcal{X} , \mathcal{Y} \right) \right)
\label{equ9}
\end{equation}
\begin{equation}
\begin{aligned}
{\mu} \left( x , \mathcal{X} , \mathcal{Y} \right) & \triangleq \mathbb{K}_{21} \left( \mathbb{K}_{11} + \xi^2 \mathbb{I} + \Lambda \left( \mathcal{X} \right) \right)^{-1} \mathcal{Y} \\
{\nu} \left( x , \mathcal{X} , \mathcal{Y} \right) & \triangleq \mathbb{K}_{22} + \xi^2 - \mathbb{K}_{21} \left( \mathbb{K}_{11} + \xi^2 \mathbb{I} + \Lambda \left( \mathcal{X} \right) \right)^{-1} \mathbb{K}_{12} 
\end{aligned} \label{equ9-2}
\end{equation}

Heteroscedasticity comes from the choice of $\Lambda$: if it is defined to be a constant function, the model becomes homoscedastic. Some papers, such as \cite{Kersting2007}, suggest modelling $\Lambda$ as another GP regression on the logarithms of the residues. This approach assumes that such logarithms follow Gaussian distributions so symmetric, and always underestimates variances by the Jensen’s inequality applied to the concave log function: the average of logarithms is always smaller than or equal to the logarithm of average!

Here, we use a more intuitive and direct form of $\Lambda$ inspired by Nadaraya-Watson kernel regression \cite{NADARAYA1964,WATSON1964,Bierens1994,LANGRENE2019} as follows: let ${\mu}_{n} \triangleq \mu\left(x_n,\mathcal{X},\mathcal{Y}\right)$ and ${\nu}_{n} \triangleq \nu\left(x_n,\mathcal{X},\mathcal{Y}\right)$ as abbreviations.
\begin{equation}
\Lambda \left( x \right) \triangleq \frac{\sum_{n=1}^{\rm N}{\left( \left(y_n-{\mu}_{n} \right)^2 + {\nu}_{n} \right)\mathcal{K}_h\left(x-x_n\right)}}{\sum_{n=1}^{\rm N}{\mathcal{K}_h\left(x-x_n\right)}} \label{equ10}
\end{equation}
, where $\mathcal{K}$ is a density kernel and $h$ is a tuning parameter which is called the bandwidth. I.e., $\Lambda\left(x\right)$ is defined as a weighted average of squares of residues, where weights are determined by how much the query SST $x$ is departed from the data. (\ref{equ10}) is derived from the following expectation over $\left. \beta \middle| \mathcal{X}, \mathcal{Y} \right.$:
\begin{equation}
\mathbb{E}_{\left. \beta \middle| \mathcal{X}, \mathcal{Y} \right.} {\left[ \frac{\sum_{n=1}^{\rm N}{\left(y_n-{\beta}_{n} \right)^2 \mathcal{K}_h\left(x-x_n\right)}}{\sum_{n=1}^{\rm N}{\mathcal{K}_h\left(x-x_n\right)}} \right]} \label{equ10-2}
\end{equation}

The choice of $\mathcal{K}$ does not substantially affect to the regression model, but the model does substantially depend on the value of $h$. One suggestion is to adapt the K-nearest neighbor bandwidth \cite{loftsgaarden1965,terrell1992}.

A GP with arcsine kernel in (\ref{equ11}) is interpretable as a one-layer neural network with infinite number of marginalized hidden nodes with a sigmoid function as the activation \cite{WILLIAMS1998}. Now we have a scalar $\eta$ and a square matrix $\Sigma$ as hyperparameters to be tuned. To avoid overfitting, it is common to assume in addition that $\Sigma$ is a diagonal matrix. Tuning hyperparameters can be done by maximizing the marginal likelihood (ML) $p\left(\mathcal{Y}\middle|\mathcal{X}\right)$ from (\ref{equ1}) and (\ref{equ2}) or by the leave-one-out cross-validation (LOO-CV): more details can be found in \cite{Rasmussen2005}.
\begin{equation}
k_{\rm{NN}} \left( x,\tilde{x} \right) \triangleq \eta^2\sin^{-1}{\left(\frac{2{f}^\mathbb{T}\Sigma \tilde{f}}{\sqrt{\left(1+2{f}^\mathbb{T}\Sigma f\right)\left(1+2{\tilde{f}}^\mathbb{T}\Sigma \tilde{f}\right)}}\right)}
\label{equ11}
\end{equation}
, where $f \triangleq \mathcal{F}\left(x\right)$ and $\tilde{f} \triangleq \mathcal{F}\left(\tilde{x}\right)$.

In our model, outliers are represented as a hidden variable $\mathcal{H}=\left\{\rm{H}_n\right\}_{n=1}^{\rm N}$, where $\rm{H}_n=0$ if $y_n$ is an inlier and 1 an outlier. $\rm{H}_n$'s are assumed to be independent and each has the following prior distribution:
\begin{equation}
p \left( {\rm{H}_n} \right) = {\rm{Bernoulli}} \left( {\rm{H}_n} \middle| q \right) \label{equ12}
\end{equation}
, where $q\in\left(0,1\right)$ is a hyperparameter not to be learned. A small $q$ implies that most of the observations are believed not to be outliers. This reflects a point of view that outliers are in essence subjective.

Once $\mathcal{H}$ is given, we specified each ${\rm{U}}_{37}^{\rm{K}'}$ observation in $\mathcal{Y}$ as follows: let $\Lambda_n \triangleq \Lambda \left( x_n \right)$ as abbreviation.
\begin{equation}
\begin{aligned}
p\left(y_n\middle| x_n,\rm{H}_n = 0 \right) & \triangleq  \mathcal{N}\left(y_n\middle|{\mu}_n,{\nu}_n + \Lambda_n \right)
\\ p\left(y_n\middle| x_n,\rm{H}_n = 1 \right) & \triangleq \frac{1}{2} \mathcal{N}\left(y_n\middle|{\mu}_n + d \sqrt{{\nu}_n + \Lambda_n},{\nu}_n + \Lambda_n \right) + \frac{1}{2} \mathcal{N}\left(y_n\middle|{\mu}_n - d \sqrt{{\nu}_n + \Lambda_n},{\nu}_n + \Lambda_n \right)
\end{aligned} \label{equ13}
\end{equation}
, where $d>0$ is a hyperparameter for how much outliers are deviated from the inlier distribution. Note that the above outlier classification is working because outputs are defined on the one-dimensional space in this problem.

By considering (\ref{equ12}) as the prior and (\ref{equ13}) as the likelihoods given ${\rm{H}}_n$, respectively, we derive the posterior distribution of ${\rm{H}}_n$ given $x_n$ and $y_n$ by Bayes' rule:
\begin{equation}
p\left(\rm{H}_n\middle| x_n,y_n\right)\propto p\left(\rm{H}_n\right)p\left(y_n\middle| x_n,\rm{H}_n\right) \label{equ14}
\end{equation}

Now we are prepared. Algorithm 1 summarizes the learning procedure of our heteroscedastic GP regression (HGPR) on ${\rm{U}}_{37}^{\rm{K}'}$ over SSTs.

\begin{algorithm}
\SetAlgoLined
\DontPrintSemicolon
 \textbf{initialize} hyperparameters and $\mathcal{H} \equiv 0$.\;
 \While{convergence}{
  \textbf{tune} kernel hyperparameters in (\ref{equ2}) and (\ref{equ11}) with $\mathcal{X},\mathcal{Y}\left|\mathcal{H}=0\right.$.\;
  \textbf{compute} $\mu$ and $\nu$ in (\ref{equ9-2}) with $\mathcal{X},\mathcal{Y}\left|\mathcal{H}=0\right.$.\;
  \textbf{choose} the bandwidths in (\ref{equ10}).\;
  \textbf{update} $\Lambda$ in (\ref{equ10}) with $\mathcal{X},\mathcal{Y}\left|\mathcal{H}=0\right.$.\;
  \textbf{sample} $\left. \mathcal{H} \middle| \mathcal{X},\mathcal{Y} \right.$ by (\ref{equ14}).\;
 }
 \textbf{return} $\mu$, $\nu$ and $\Lambda$.
 \caption{HGPR on ${\rm{U}}_{37}^{\rm{K}'}$ over SSTs}
\end{algorithm} \label{algorithm1}

One possible way of utilizing the obtained GP regression model in (\ref{equ9}) is plugging it in the following Bayesian inversion:
\begin{equation}
p\left( \tilde{x} \middle| \tilde{y} \right) \propto p\left( \tilde{y} \middle| \tilde{x} \right) p\left(\tilde{x}\right) \label{equ14-2}
\end{equation}
, where $\tilde{y}$ is the observed ${\rm{U}}_{37}^{\rm{K}'}$ data and $\tilde{x}$ is the associated SSTs to be inferred. The prior $p\left(\tilde{x}\right)$ can be given as a distribution of SSTs given their geographical information, for instance.

In general, (\ref{equ14-2}) does not have a closed form. Because SSTs are of one dimension, applying the Markov-chain Monte Carlo (MCMC) is enough to sample from the posterior $p\left( \tilde{x} \middle| \tilde{y} \right)$; if the event to infer is of high dimension, a variational inference to approximate the posterior with a known distribution must be considered.

\section{Data}\label{sec3}

We used the dataset same with \cite{TIERNEY2018}, after discarding those in the locations that it excluded from their analysis. The rest of data with 1274 observations were used. We could discard more from them but did not do so for checking whether or not our model could classify apparent outliers desirably. Details about data are in \cite{TIERNEY2018}.

For the density kernel and bandwidth in (\ref{equ10}), we adapted a Gaussian kernel having an unbounded support and the K-nearest neighbor bandwidth where $\rm{K}$ is selected by the LOO-CV among 10 candidates, from 1\% to 10\% of the data per 10 iterations. Hyperparameters $q$ in (\ref{equ12}) and $d$ in (\ref{equ13}) were set to be 0.065 and 2.48, respectively: these values lead the marginal likelihood of $y_n$ from (\ref{equ12}) and (\ref{equ13}) to approximating the Student's t-distribution with 6 as the degree of freedom.

We also regularized the raw data $\mathcal{X}$ and $\mathcal{Y}$ by $x \leftarrow x / 6.5656 - 3.0205$ and $y \leftarrow y / 0.2104 - 3.3642$ so that they fit to the Student's t-distribution with 6 as the degree of freedom, and then took a feature function $\mathcal{F} \left( x \right) = (1,x)^{\mathbb{T}}$ on the regularized $\mathcal{X}$. To identify and regularize the model, we defined $\Sigma$ in (\ref{equ11}) to be a diagonal matrix, where the first entry is fixed to be 1.

\section{Results}\label{sec4}

\begin{figure}
\centering
\includegraphics[width=1.0\textwidth]{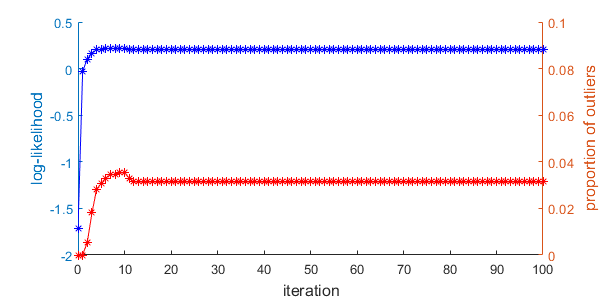}
\caption{Average log-likelihoods of inliers and proportions of outliers over iterations.}
\label{fig1}
\end{figure}

\begin{figure}
\centering
\includegraphics[width=1.0\textwidth]{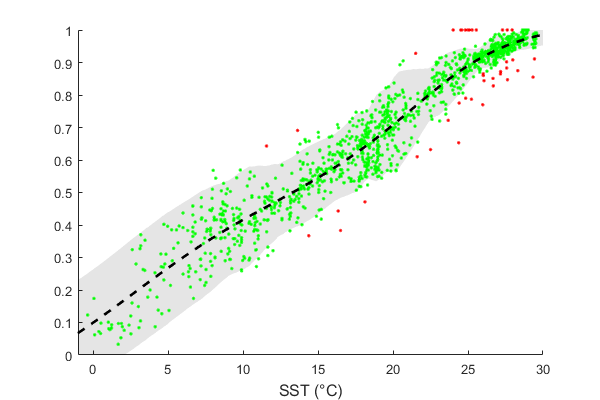}
\caption{The learned regression over SSTs. Green and red points are inliers and outliers, respectively. The dashed line shows the inferred means over SSTs and shaded region is the 95\% confidence band.}
\label{fig2-1}
\end{figure}

\begin{figure}
\centering
\includegraphics[width=1.0\textwidth]{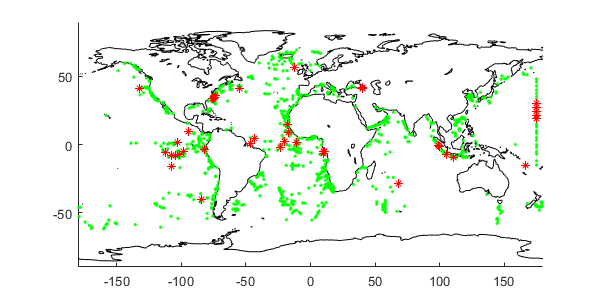}
\caption{The world map with data locations. Green and red points are inliers and outliers, respectively.}
\label{fig2-2}
\end{figure}

\begin{figure}
\centering
\includegraphics[width=1.0\textwidth]{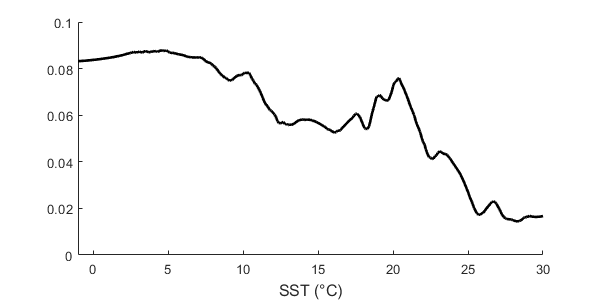}
\caption{Inferred standard deviations over SSTs.}
\label{fig4-1}
\end{figure}

\begin{figure}
\centering
\includegraphics[width=1.0\textwidth]{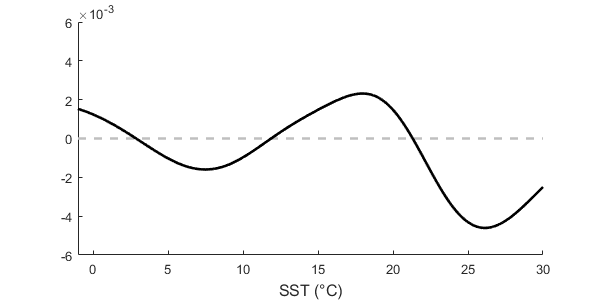}
\caption{Inferred curvatures over SSTs.}
\label{fig4-2}
\end{figure}

\begin{figure}
\centering
\includegraphics[width=1.0\textwidth]{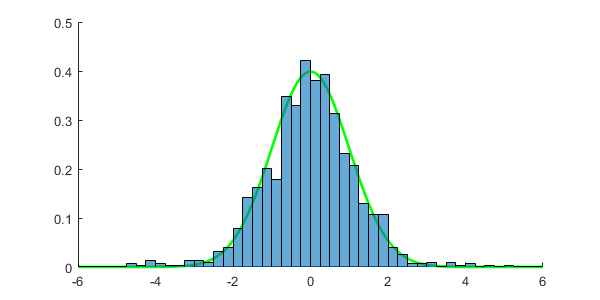}
\caption{The histogram of standardized residuals. The green graph is the standard normal distribution.}
\label{fig3-1}
\end{figure}

\begin{figure}
\centering
\includegraphics[width=1.0\textwidth]{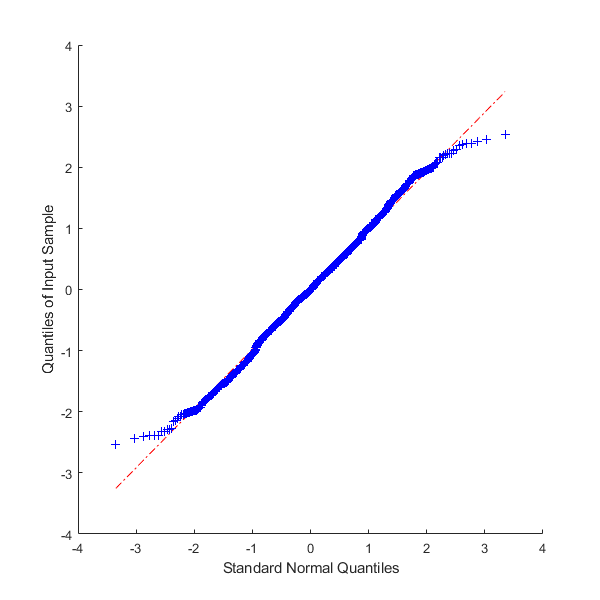}
\caption{The Q-Q plot of inliers.}
\label{fig3-2}
\end{figure}

\begin{figure}
\centering
\includegraphics[width=1.0\textwidth]{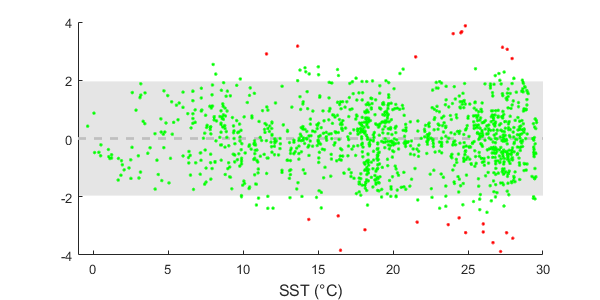}
\caption{A plot of standardized residuals over SSTs. Green and red points are inliers and outliers, respectively. The shaded region covers the 95\% confidence intervals $\left[ -1.96,1.96 \right]$.}
\label{fig3-3}
\end{figure}

After 100 iterations in about 7 minutes, we obtained the results shown in the eight figures, from \ref{fig1} to \ref{fig3-3}. Finally selected $\rm{K}$ was 4\%. Figure \ref{fig1} shows the average log-likelihoods of inliers (blue) and proportions of inferred outliers (red). It shows convergence of the learning procedure after around the ${15}^{\rm{th}}$ iteration. Only about 3\% of the data were classifed as outliers. Learned kernel hyperparameters are the following:
\begin{equation}
\begin{aligned}
\eta & = 2.1962 \\
\Sigma & = 
\begin{bmatrix}
1 & 0 \\
0 & 0.4712^2 \\
\end{bmatrix} \\
\xi & = 2.7253 \times 10^{-12}
\end{aligned} \label{equ15-0}
\end{equation}

Figure \ref{fig2-1} represents the inferred regression on SSTs with the classification of inliers (green) and outliers (red), and figure \ref{fig2-2} those on the world map. It clearly shows the heteroscedasticity of observations and the associated model, as figure \ref{fig4-1}. Also, apparent outliers at 22-26$\degree \rm{C}$ of the upper boundary of the plot are classified as so. Classified outliers above $25 \degree \rm{C}$ below the model, however, could be from another cluster. This can be adjusted by tuning hyperparameters in section \ref{sec3}. One advantage of GP regression is that the inferred mean at each input is expressed in a closed form and differentiable as much as the adopted kernel. Curvatures of the inferred means are shown in figure \ref{fig4-2}. \cite{TIERNEY2018} suggests that slopes of the means start to change at 23.4$\degree \rm{C}$, which is roughly consistent with our inference, but some more changes are also captured in the inferred model.

To visually check how residuals are treated appropriately by our heteroscedastic model, figures \ref{fig3-1} to \ref{fig3-3} were also plotted. A standardized residual $r_n$ at $x_n$ is defined as follows:
\begin{equation}
r_n \triangleq \frac{y_n - \mu_n}{\sqrt{\nu_n + \Lambda_n}}
\label{equ15}
\end{equation}

I.e., if means and variances of the GP regression model are properly inferred, standardized residuals at inliers must follow the standard normal distribution.

In figure \ref{fig3-1}, most of the standardized residuals seem to follow the standard normal distribution, and figure \ref{fig3-2} supports that assertion as the Q-Q plot of inliers. In addition, figure \ref{fig3-3} shows that standardized residuals are barely correlated with SSTs: the correlation between the pairs classified as inliers is $0.0019$. These results strongly suggest that our GP regression model appropriately infer the heteroscedasticity of the model.

\section{Conclusion}\label{sec5}

Our GP regression on ${\rm{U}}_{37}^{\rm{K}'}$ appropriately explains the heteroscedasticity of data and provides a probabilistic model with explicit distributions. It converges quickly, in about 7 minutes, even if we do not assume any informative prior structure (see $\vec{0}$ in (\ref{equ2})) on the regression. Thus, this work can be considered as a success of the nonparametric approach on the real data.

However, a GP has several disadvantages to overcome. Firstly, it requires at least one matrix inversion in constructing distributions, where its size is the same as the number of observations. There are only 1274 data, which is affordable in the currently available computing power, so it was not problematic in this case. We used two-dimensional features, which do not suffer from the curse of dimensionality.

Nonetheless, in this paper a model based on GPs shows its effectiveness on explaining ${\rm{U}}_{37}^{\rm{K}'}$ observations over SSTs, where relatively moderate amount of data are given and inputs are of low-dimensional. One more advantage of the nonparametric approaches is that it does not depend much on the specific characteristics of data: it is possible to adapt the same approach to any data to do regression.

Codes which run on MATLAB can be found in \url{https://github.com/eilion/HGPR\_SST\_Proxy\_Cal}.

\section{Acknowledgements}\label{sec6}

This paper is based on the works supported by the Division of Applied Mathematics in Brown University, by the National Science Foundation (NSF) under a grant number OCE-1760838, and by the Kwanjeong Educational Foundation.

\bibliographystyle{unsrt}
\bibliography{reference}

\end{document}